\documentstyle[sprocl,psfig]{article}
\def\NPB{{\em Nucl. Phys.} B}
\def\PLB{{\em Phys. Lett.}  B}
\def\PRL{\em Phys. Rev. Lett.} 
\def\PRD{{\em Phys. Rev.} D}
\def\ZPC{{\em Z. Phys.} C}
\def\st{\scriptstyle}

\def\be{\begin{equation}}
\def\ee{\end{equation}}
\def\bea{\begin{eqnarray}}
\def\eea{\end{eqnarray}}

\newcommand{\appendixA}{\setcounter{equation}{0}
\def\theequation{\rm{A}.\arabic{equation}}\section*}
\newcommand{\appendixB}{\setcounter{equation}{0}
\def\theequation{\rm{B}.\arabic{equation}}\section*}

\catcode`\@=11
\def\marginnote#1{}
\newcount\hour
\newcount\minute
\newtoks\amorpm
\hour=\time\divide\hour by60
\minute=\time{\multiply\hour by60 \global\advance\minute by-
\hour}
\edef\standardtime{{\ifnum\hour<12 \global\amorpm={am}%
    \else\global\amorpm={pm}\advance\hour by-12 \fi
    \ifnum\hour=0 \hour=12 \fi
    \number\hour:\ifnum\minute<100\fi\number\minute\the\amorpm}}
\edef\militarytime{\number\hour:\ifnum\minute<100\fi\number\minute}
\def\draftlabel#1{{\@bsphack\if@filesw {\let\thepage\relax
  \xdef\@gtempa{\write\@auxout{\string
    \newlabel{#1}{{\@currentlabel}{\thepage}}}}}\@gtempa
    \if@nobreak \ifvmode\nobreak\fi\fi\fi\@esphack}
     \gdef\@eqnlabel{#1}}
\def\@eqnlabel{}
\def\@vacuum{}
\def\draftmarginnote#1{\marginpar{\raggedright\scriptsize\tt#1}}
\def\draft{\oddsidemargin -.5truein
        \def\@oddfoot{\sl preliminary draft \hfil
        \rm\thepage\hfil\sl\today\quad\militarytime}
        \let\@evenfoot\@oddfoot \overfullrule 3pt
        \let\label=\draftlabel
        \let\marginnote=\draftmarginnote
   
\def\@eqnnum{(\theequation)\rlap{\kern\marginparsep\tt\@eqnlabel}%
\global\let\@eqnlabel\@vacuum}  }
\def\preprint{\twocolumn\sloppy\flushbottom\parindent 1em
        \leftmargini 2em\leftmarginv .5em\leftmarginvi .5em
        \oddsidemargin -.5in    \evensidemargin -.5in
        \columnsep 15mm \footheight 0pt
        \textwidth 250mmin      \topmargin  -.4in
        \headheight 12pt \topskip .4in
        \textheight 175mm
        \footskip 0pt
        
\def\@oddhead{\thepage\hfil\addtocounter{page}{1}\thepage}
        \let\@evenhead\@oddhead \def\@oddfoot{} \def\@evenfoot{} 
}
\def\titlepage{\@restonecolfalse\if@twocolumn\@restonecoltrue\onecolumn
     \else \newpage \fi \thispagestyle{empty}\c@page\z@
        \def\thefootnote{\fnsymbol{footnote}} }
\def\endtitlepage{\if@restonecol\twocolumn \else  \fi
        \def\thefootnote{\arabic{footnote}}
        \setcounter{footnote}{0}}  
\catcode`@=12
\relax
\def\be{\begin{equation}}
\def\ee{\end{equation}}
\def\bea{\begin{eqnarray}}
\def\eea{\end{eqnarray}}
\def\simlt{\stackrel{<}{{}_\sim}}
\def\simgt{\stackrel{>}{{}_\sim}}

\def\NPB#1#2#3{{\it Nucl.~Phys.} {\bf{B#1}} (19#2) #3}
\def\PLB#1#2#3{{\it Phys.~Lett.} {\bf{B#1}} (19#2) #3}
\def\PRD#1#2#3{{\it Phys.~Rev.} {\bf{D#1}} (19#2) #3}
\def\PRL#1#2#3{{\it Phys.~Rev.~Lett.} {\bf{#1}} (19#2) #3}
\def\ZPC#1#2#3{{\it Z.~Phys.} {\bf C#1} (19#2) #3}
\def\PTP#1#2#3{{\it Prog.~Theor.~Phys.} {\bf#1}  (19#2) #3}

\def\PR#1#2#3{{\it Phys.~Rep.} {\bf#1} (19#2) #3}
\def\RMP#1#2#3{{\it Rev.~Mod.~Phys.} {\bf#1} (19#2) #3}
\def\HPA#1#2#3{{\it Helv.~Phys.~Acta} {\bf#1} (19#2) #3}

\def\mst11{m_{\;\widetilde{t}_{1}}}
\def\msti{m_{\;\widetilde{t}_i}}
\def\mstj{m_{\;\widetilde{t}_j}}
\def\msbi{m_{\;\widetilde{b}_i}}
\def\msbj{m_{\;\widetilde{b}_j}}
\def\st{\;\widetilde{t}}
\def\sb{\;\widetilde{b}}
\def\mst22{m_{\;\widetilde{t}_{2}}}
\def\mst12{m_{\;\widetilde{t}_{1,2}}}

\def\msb11{m_{\;\widetilde{b}_{1}}}
\def\msb22{m_{\;\widetilde{b}_{2}}}
\def\msb12{m_{\;\widetilde{b}_{1,2}}}

\def\mtilde2{\widetilde{m}^{2}}

\def\exis{\varphi}
\relax
\begin{document}

\begin{flushright}
IEM-FT-153/97 \\
hep--ph/9703412\\
\end{flushright}

\vspace{0.5cm}
\title{CONSTRAINTS ON THE HIGGS BOSON PROPERTIES FROM THE EFFECTIVE
POTENTIAL~\footnote{To appear in {\it Perspectives on Higgs Physics II},
Ed. G.L.~Kane, World Scientific, Singapore.}}
\author{ M. QUIROS }
\address{Instituto de Estructura de la Materia (CSIC), \\
Serrano 123, 28006-Madrid, SPAIN}
\maketitle\abstracts{
We review the constraints on Higgs boson properties from
effective potential methods. In the Standard Model, the 
requirement of stability (or metastability) of the standard
electroweak minimum puts an upper bound on the scale of 
new physics as a function of the Higgs mass. This upper bound
is below the Planck scale if the Higgs weights $\simlt$ 130 GeV. 
In supersymmetric extensions of the Standard Model the former 
methods are useful to compute the Higgs mass spectrum, and couplings, 
after resumming higher loop effects. In particular, if the Higgs mass 
weights $\simgt$ 130 GeV, the minimal supersymmetric extension of the 
Standard Model will be ruled out, while its non-minimal supersymmetric 
extension containing singlets is only marginally allowed.}

\section{Introduction}

The effective potential~\cite{CW}$^-$\cite{Ford} 
(the effective action at zero external
momentum) $V_{\rm eff}$ is a very useful tool to compute the
vacuum of a quantum field theory, and thus to determine its mass
spectrum. The effective potential admits a loop expansion as:
\be
\label{expansion}
V_{\rm eff}=\sum_{n=0}^{\infty} V_n
\ee
and its calculation involves (at $n\geq1$) a mass parameter
$\mu$ upon which it explicitly depends. For instance in
mass-independent renormalizations schemes, which are those usually
employed, as $\overline{\rm
MS}$~\cite{MS}, or $\overline{\rm DR}$~\cite{DR}, the one-loop
effective potential can be written in the 't Hooft-Landau gauge
as
\be
\label{oneloop}
V_1=\frac{1}{64\pi^2}\ {\rm Str}\;
m^4\left[\log\frac{m^2}{\mu^2}-C\right],  
\ee
where the supertrace operator counts positively (negatively) 
the number of degrees of freedom for the different bosonic (fermionic) 
fields, $C$ is a (constant) diagonal matrix which
depends on the renormalization scheme, and $\mu$ is the
renormalization group scale on which all dimensionless and
dimensional couplings of the theory depend.

Being $\mu$ an unphysical scale, the effective
potential should not depend on the choice of it. This property
can be achieved if a change in the scale $\mu$ can be
compensated by appropriate changes in the couplings and field
rescalings. Mathematically it can be formulated by a
renormalization group equation (RGE) satisfied by the effective
potential, which admits as solution:
\be
\label{improved}
V_{\rm eff}\equiv V_{\rm eff}(\mu(t),\lambda_i(t);\phi_a(t))
\ee
where $\lambda_i(t)$ are all the (running) parameters of the theory,
the scale $\mu(t)$ is related to the running parameter $t$ by
\be
\label{scale}
\mu(t)=\mu e^{t}
\ee
and the running (Higgs) fields are $(a=1,\dots , N)$
\be
\label{fields}
\phi_a(t)=\xi_{a}(t)\phi_a\equiv 
e^{-\int_0^t \gamma_a(t') dt'}\ \phi_a
\ee
where $\phi_a$ are the classical fields on which the generating
functional of 1PI Green functions depend, and $\gamma_{a}(t)$ are
the corresponding anomalous dimensions. We are assuming here
that the fields $\phi_a$ have different quantum numbers and so
they are not mixed by radiative corrections. Of course this
situation is not the most general one, but it corresponds to the
cases that will be studied in this paper: the Standard Model
(SM), with a single Higgs doublet, the Minimal
Supersymmetric Standard Model (MSSM), which contains two Higgs doublets
with opposite hypercharges, and its extension with one singlet superfield.

Now the complete effective potential (\ref{improved}) and its
$n^{\rm th}$ derivative matrix, defined by
\be
\label{nderivada}
V_{\rm eff}^{a_1\dots a_n}=\xi_{a_1}(t)\dots\xi_{a_n}(t)
\frac{\partial^n}{\partial\phi_{a_1}(t)\dots
\partial\phi_{a_n}(t)} V_{\rm eff}
\ee
satisfies the scale-independence condition
\be
\label{independencia}
\frac{d\;V_{\rm eff}^{a_1\dots a_n}}{dt}=0
\ee
By defining the running masses by
\be
\label{runmass}
\overline{m}_{ab}^2(t)=\left.\frac{\partial^2}{\partial\phi_a(t) 
\partial\phi_b(t)} V_{\rm eff}\right|_{\langle\phi(t)
\rangle},
\ee
and using the scale independence of the effective potential
(\ref{independencia}), we obtain the scale variation:
\be
\label{varmasses}
\frac{d\; \overline{m}_{ab}^2(t)}{dt}=\left[\gamma_a(t)+\gamma_b(t)\right] 
\overline{m}^2_{ab}(t)
\ee

Of course the running masses are not physical quantities since
they depend on the renormalization scale as in
(\ref{varmasses}), and also on the gauge and renormalization
scheme, as the effective potential does. 
However, one can compute, from
the running masses the physical (pole) masses as follows.
The inverse of the renormalized propagator with momentum $p$ 
can be written~\footnote{Since we are
mainly interested in masses (not widths) we understand
implicitly that we are considering the real parts of the inverse
propagators and of the self-energies.} on general grounds as
\be
\label{inverseprop}
\Gamma_{ab}(p^2)=p^2\delta_{ab}-[m^2_{ab}+\Pi_{ab}(p^2)]
\ee
where $m^2_{ab}$ are the renormalized masses and $\Pi_{ab}(p^2)$
are the $\overline{\rm MS}$, or $\overline{\rm DR}$-renormalized
self energies, obtained by subtracting from the corresponding
unrenormalized quantities (regulated via dimensional
regularization) the divergent terms proportional to 
$1/\epsilon-\gamma_E+\log 4\pi$, where the space-time
dimensionality is written as $D=4-2\epsilon$ and $\gamma_E$ is
the Euler constant.

The relation between the running mass $\overline{m}_{ab}$ and
the inverse propagator,
\be
\label{relacionprop}
\Gamma_{ab}(0)=-\overline{m}_{ab}^2(t)
\ee
allows to write the relation between the renormalized and the
running masses, as
\be
\label{relacionmasa}
\overline{m}_{ab}^2=m_{ab}^2+\Pi_{ab}(0) 
\ee
and hence the inverse propagator is usually written as
\be
\label{inverseprop2}
\Gamma_{ab}(p^2)=p^2\delta_{ab}-[\overline{m}^2_{ab}+
\Delta\Pi_{ab}(p^2)]
\ee
\be
\label{deltapi}
\Delta\Pi_{ab}(p^2)\equiv \Pi_{ab}(p^2)-\Pi_{ab}(0)
\ee
The physical masses are then defined as the poles of the
diagonalized inverse propagator (\ref{inverseprop2}), or the
solution to the equation,
\be
\label{pole}
\det\left[p^2 \delta_{ab}-(m^2_{ab}+\Pi_{ab}(p^2)\right]=0
\ee
or, equivalently,
\be
\label{polerun}
\det\left[p^2\delta_{ab}-(\overline{m}_{ab}^2+\Delta\Pi_{ab}(p^2)\right]=0
\ee

In a diagrammatic calculation the physical masses are obtained as
solutions of Eq. (\ref{pole}). Then, after computing the matrix
$m_{ab}+\Pi_{ab}(p^2)$ and finding its eigenvalues $\mu_A^2(p^2)$,
($A=1,\dots ,N$), the physical (pole) masses $M_A$ are 
the solutions of the equations 
\be
\label{masaspolo}
M_A^2=\mu_A^2(M_A^2)
\ee
which is obviously the solution to Eq. (\ref{pole}).
Of course, Eqs. (\ref{pole}) and (\ref{polerun}) are
completely equivalent and both have the same solution. Obviously
diagrammatic methods, solving Eq. (\ref{pole}), and 
effective potential methods, solving Eq. (\ref{polerun}), yield
the same result, at they should. However effective potential
methods have the advantage that $\overline{m}_{ab}^2$ 
already contains the dominant
part of the one-loop corrections and Eq.~(\ref{polerun})
can be most easily solved approximately. 
The way of proceeding is by diagonalization of the matrix
of running masses $\overline{m}_{ab}^2$ 
with eigenvalues $\overline{m}_A^2$, 
$\phi_A$ being the new mass eigenstate fields. The physical masses
$M_A$ are then given by,
\be
\label{masaspolo2}
M_A^2=\overline{m}_A^2+\Delta\Pi_{AA}(M_A^2)
\ee
where the self-energies in (\ref{masaspolo2}) are computed in
the mass eigenstate basis $\phi_A$.

Let us notice that, unlike the running masses, $\overline{m}_A$,
the pole masses $M_A$ do not depend on the renormalization
scale, this fact being guaranteed by the invariance of the
effective action with respect to the scale, and hence by the
renormalizability of the theory. In practical calculations this
invariance can be considered as a cross-check of the result.
For the case of the SM this has been explicitly shown~\cite{CEQR}.

The contents of this article are as follows. In Section 2
the effective potential methods will be applied to the Standard
Model. In particular lower stability (and 
metastability) bounds and upper perturbativity
bounds will be obtained for the pole Higgs mass. Section 3 will be
devoted to the MSSM, where the radiatively corrected spectrum of
Higgs masses and couplings will be deduced from effective potential
methods. Section 4 will contain results on upper bounds on the lightest
Higgs boson mass in extensions of the MSSM: in particular extensions
containing Higgs singlets, where the properties of gauge coupling
constant unification observed at LEP are not spoiled. Section 5 tries
to answer the question on, what if a light Higgs (with a mass below
$\sim$ 130 GeV) is discovered at LEP2 or FNAL? Finally Section 6
contains some general conclusions. The self-energies, $\Delta\Pi_{ab}$
relating running and physical masses for the cases of the Standard Model
and MSSM are listed in Appendices A and B, respectively.

\section{The Standard Model}

In the SM the Higgs boson is an SU(2)$_L$ doublet with hypercharge 1/2 
given by:
\be
\label{higgsSM}
H=\left(
\begin{array}{c}
\chi^+ \\
{\displaystyle \frac{\phi+i\chi^0}{\sqrt{2}} }
\end{array}
\right)
\ee
where $\phi$ is the physical Higgs, that must acquire a vacuum 
expectation value (VEV) $v=(\sqrt{2} G_F)^{-1/2}$. The fields $\chi^{\pm 
0}$ are Goldstone bosons, which become massless in the vacuum of the 
theory. Therefore, the effective potential for the SM depends on a single 
background field $\phi$.

The renormalization group improved effective potential of the SM, $V$, can
be written in the 't Hooft-Landau gauge and the $\overline{MS}$ 
scheme as~\cite{Einhorn}
\be
\label{veff}
V[\mu(t),\lambda_i(t);\phi(t)]\equiv V_0 + V_1 + \cdots \;\; ,
\ee
where $\lambda_i\equiv (g,g',\lambda,h_t,m^2)$ runs over all
dimensionless and dimensionful couplings and $V_0$, $V_1$ are 
the tree level potential and the one-loop correction, respectively, namely
\be
\label{v0}
V_0=-{\displaystyle\frac{1}{2}}m^2(t)\phi^2(t) +
{\displaystyle\frac{1}{8}}\lambda(t)\phi^4(t),
\ee
\be
\label{v1}
V_1={\displaystyle\sum_{i=1}^5}{\displaystyle\frac{n_i}{64\pi^2}}
m_i^4(\phi)\left[\log{\displaystyle\frac{m_i^2(\phi)}{\mu^2(t)}}
-c_i\right]+\Omega(t),
\ee
$\Omega(t)$ being
the one-loop contribution to the cosmological constant~\cite{Einhorn},
and
\be
m_i^2(\phi)=\kappa_i\phi^2(t)-\kappa_i',
\label{masas}
\ee
with non-zero coefficients $
n_W=6,\; \kappa_W=g^2(t)/4,\; c_W=5/6;
\; n_Z=3,\ \kappa_Z=[g^2(t)+g'^2(t)]/4,\; c_Z=5/6;
\; n_t=-12,\ \kappa_t=h_t^2(t)/2,\; c_t=3/2;
\; n_{\phi}=1,\  \kappa_{\phi}=3\lambda(t)/2,
\; \kappa_{\phi}'=m^2(t),
\; c_{\phi}=3/2;
\; n_{\chi}=3,\  \kappa_{\chi}=\lambda(t)/2,
\; \kappa_{\chi}'=m^2(t),
\; c_{\chi}=3/2.$

It has been shown~\cite{BKMN} that the L-loop  effective potential
improved by (L+1)-loop RGE resums all Lth-to-leading logarithm contributions.
Consequently, we will consider all the $\beta$- and $\gamma$-functions of
the previous parameters to two-loop order, so that our calculation will
be valid up to next-to-leading logarithm approximation.

As has been pointed out~\cite{Einhorn}, working with $\partial V/
\partial \phi$ (and higher derivatives) rather than with $V$ itself allows
to ignore the cosmological constant term $\Omega$. In fact,
the structure of the potential
can be well established once we have determined the values of $\phi$, say
$\phi_{\rm ext}$, at which $V$ has extremals (maxima or minima), thus we only
need to evaluate $\partial V/
\partial \phi$ and $\partial^2 V/
\partial \phi^2$. From Eq.~(\ref{veff}), the value of $\phi_{\rm ext}$
is given by
\be
\label{first}
\left.\frac{\partial V}{\partial \phi(t)}\right|_{\phi(t)=
\phi_{\rm ext}(t)}=0
\;\;\Rightarrow\;\; \phi^2_{\rm ext}(t)=\frac{2m^2
+{\displaystyle\sum_i\frac{n_i\kappa_i\kappa'_i}{8\pi^2}}
\left[\log\frac{m_i^2(\phi)}{\mu^2(t)} -c_i+\frac{1}{2}
\right]}{\lambda+{\displaystyle\sum_i\frac{n_i\kappa_i^2}{8\pi^2}}
\left[\log\frac{m_i^2(\phi)}{\mu^2(t)} -c_i+\frac{1}{2}
\right]} .
\ee
The second derivative is given by
\be
\label{second}
\left.\frac{\partial^2 V}{\partial \phi^2(t)}\right|_{\phi(t)=
\phi_{\rm ext}(t)}
=2m^2+\sum_i\frac{n_i\kappa_i\kappa'_i}{8\pi^2}
\left[\log\frac{m_i^2(\phi)}{\mu^2(t)} -c_i+\frac{1}{2}
\right] +\phi^2(t)\sum_i\frac{n_i\kappa_i^2}{8\pi^2},
\ee
where we have used (\ref{first}). Eq. (\ref{second}) can be written
in a more suggestive form as
\bea
\label{secondbis}
\left.\frac{\partial^2 V}{\partial \phi^2(t)}\right|_{\phi(t)=
\phi_{\rm ext}(t)}
&=&2m^2+\frac{1}{2}(\beta_{\lambda}-4\gamma\lambda)\phi^2(t)
\vspace{0.3cm}\nonumber\\
&+&\frac{1}{2}(\beta_{m^2}-2\gamma m^2)\left[\sum_{i,\kappa_i'\neq 0}
\log\frac{m_i^2(\phi)}{\mu^2(t)} -2 \right] ,
\eea
where $\beta_{\lambda}$, $\beta_{m^2}$ and $\gamma$ are one-loop
$\beta$- and $\gamma$-functions.

We would like to stress the fact that even though
the whole effective potential is scale-invariant, the one-loop approximation
is {\em not}. Therefore, one needs a criterion to choose the appropriate
renormalization scale in the previous equations. A sensible 
criterion~\cite{CEQR} is to choose the scale, say $\mu^*=\mu(t^*)$, where the
effective potential is more scale-independent. $\mu^*$ turns out to be a
certain average of the $m_i(\phi)$ masses. Actually, it was shown~\cite{CEQR} 
that any choice of $\mu^*$
around the optimal value produces the same results for physical quantities up
to tiny differences.

The running mass $m_H^2$, defined as the
second derivative of the renormalized effective potential
(see Eqs.~(\ref{runmass}) and (\ref{varmasses})), is given by
\begin{equation}
\label{mhmh}
m_H^2=\frac{\partial^2 V_{eff}}{\partial \phi^2}=
-\Gamma_{R}(p^2=0)=
m_R^2+\Pi_R(p^2=0) \ \ .
\end{equation}
where $m_R$  is the renormalized mass, while the physical (pole)
mass satisfies the relation
\begin{equation}
\label{MHphys}
M_H^2=m_R^2+\Pi_R(p^2=M_H^2)\ \ .
\end{equation}
Comparing (\protect\ref{MHphys}) and (\protect\ref{mhmh}) we have
\begin{equation}
\label{MHdelta}
M_H^2= m_{H}^2 +\Delta \Pi,
\end{equation}
where we have defined (we drop the subscript R from $\Pi_R$)
\begin{equation}
\label{deltadepi}
\Delta \Pi\equiv\Pi (p^2=M_H^2)-\Pi (p^2=0).
\end{equation}
The complete expression for the quantity $\Delta\Pi$ is given in
appendix A.
\subsection{Stability bounds}

The solution to (\ref{first}) describes the standard electroweak (EW)
minimum, but 
might also describe other maxima and non-standard minima.
The preliminary question one should now ask is: When the standard EW 
minimum becomes
metastable, due to the appearance of a deep non-standard 
minimum? This question was
addressed in past years~\cite{L} taking into account leading-log (LL) and 
part of next-to-leading-log (NTLL) corrections. 
More recently, calculations have incorporated all NTLL
corrections~\cite{AI,CEQ} 
resummed to all-loop by the renormalization group equations,
and considered pole masses for the top-quark and the Higgs-boson. 
From the requirement of a stable (not metastable) standard EW minimum 
we obtain a lower bound on
the Higgs mass, as a function of the top mass, labelled by the values of 
the SM cutoff (stability bounds). Our
result~\cite{CEQ} is lower than previous estimates.

\begin{figure}[hbt]
\centerline{
\psfig{figure=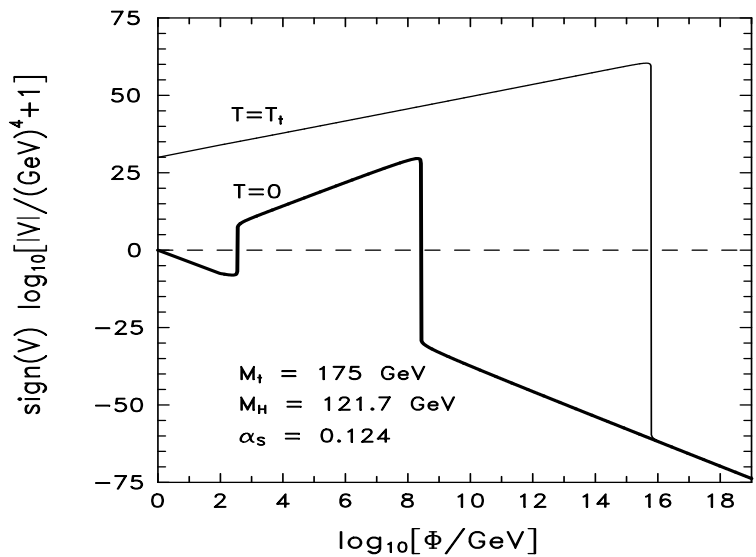,height=7.5cm,width=7cm,bbllx=4.75cm,bblly=3.cm,bburx=14.25cm,bbury=16cm}}
\caption{Plot of the effective potential for $M_t=175$ GeV, $M_H=121.7$
GeV at $T=0$ (thick solid line) and $T=T_t=2.5\times 10^{15}$ GeV 
(thin solid line).}
\label{fval1}
\end{figure}

The problem to attack is easily stated as follows.
The condition (\ref{first})
is obviously satisfied at the EW minimum where
$\langle\phi\rangle=v\sim 246$ GeV, $\lambda\sim(m_H/v)^2>1/16$,
$m^2\sim m_H^2/2$ and $V''(\langle\phi\rangle)>0$ (a minimum).
However, it can also be satisfied for values
of the field $\phi\gg v$ and, since $m={\cal O}(100)$ GeV, 
for those values
$$
\lambda\sim\left(\frac{m}{\phi}\right)^2\ll 1.
$$
Therefore, for the non-standard extremals we have 
\begin{eqnarray}
\label{minmax}
\beta_\lambda < 0 & \Longrightarrow & V''<0\ {\rm
maximum}\nonumber \\
\beta_\lambda > 0 & \Longrightarrow & V''>0\ {\rm minimum}. 
\end{eqnarray}

The one-loop effective potential of the SM improved by 
two-loop RGE has been shown to
be highly scale independent~\cite{CEQR} and, therefore, very reliable for the 
present study. 
In Fig.~\ref{fval1} we show (thick solid line) 
the shape of the effective potential for~\footnote{$M_t$ stands for the
physical (pole) top-quark mass, as opposite to the top-quark running mass
defined in (\ref{masas}).}
$M_t=175$ GeV 
and $M_H=121.7$ GeV. We see the appearance of the non-standard maximum,
$\phi_M$, while the global
non-standard minimum has been cutoff at $M_{P\ell}$. 
We can see from Fig.~\ref{fval1} the 
steep descent from the non-standard maximum. Hence, 
even if the non-standard minimum is beyond 
the SM cutoff, the
standard minimum becomes metastable and might be destabilized. So for fixed 
values of $M_H$ and
$M_t$ the condition for the standard minimum not to become metastable is 
\be
\label{condstab}
\phi_M \simgt \Lambda
\ee

\begin{figure}[hbt]
\centerline{
\psfig{figure=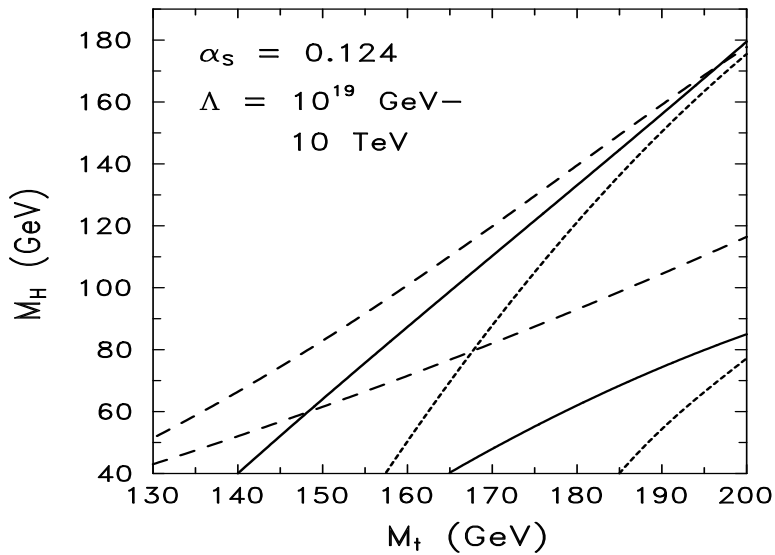,height=7.5cm,width=7cm,bbllx=5.cm,bblly=2.cm,bburx=14.5cm,bbury=15cm}}
\caption{Lower bounds on $M_H$ as a function of $M_t$, for
$\Lambda=10^{19}$ GeV (upper set) and $\Lambda=10$ TeV (lower set). 
The dashed curves
correspond to the stability bounds and  the solid (dotted) 
ones to the metastability bounds at finite (zero) temperature.}
\label{fval2}
\end{figure}

Equation (\ref{condstab}) makes the stability condition $\Lambda$-dependent. 
In fact we have plotted
in Fig.~\ref{fval2} the stability condition on $M_H$ versus $M_t$ for 
$\Lambda=
10^{19}$ GeV and 10 TeV. The stability 
region corresponds to the region above the dashed curves.
Accurate fits of the bounds on $M_H$ are,
\be
\label{cota3}
M_H[GeV] > 52+0.64 (M_t[GeV]-175)-0.50 \frac{\alpha_s(M_Z)-0.12}
{0.006}
\ee
for $\Lambda=1$ TeV, and
\be
\label{cota19}
M_H[GeV] > 133+1.92 (M_t[GeV]-175)-4.28 \frac{\alpha_s(M_Z)-0.12}
{0.006}
\ee
for $\Lambda=10^{19}$ GeV.

We want to conclude this section with a few comments about the errors
affecting our analysis~\cite{CEQ}. Errors come from: 
a) The evaluation of the top-quark pole mass $M_t$, where we have 
neglected two-loop QCD and one-loop electroweak corrections; 
b) Higher (two)-loop corrections which have been neglected in the 
effective potential; c) The fact that bounds
are computed in a particular (Landau) gauge. The gauge dependence of
the result can be translated into an error~\footnote{The gauge
dependence of the bound has recently been stressed~\cite{Willey}
in a toy model, with a U(1) gauge symmetry and fermion fields which
mimic the effect of the top quark in the Standard Model, whose effective
potential in a class of gauges parametrized by a real parameter $\xi$
($\xi=0$ in Landau gauge) has been worked out. The effective potential,
as well as the $\beta$ and $\gamma$-functions, depends on the RGE
invariant combination $\xi g^2$. These authors have pointed out that for
extremely large values of the parameter $\xi g^2$ the effect of the
one-loop correction to the effective potential on the Higgs mass bound
blows up. This is just a reflection of the failure of
perturbation theory, and higher loop corrections should affect
dramatically the results of Ref.~\cite{Willey}  for these values of  
$\xi g^2$. However, for moderate values of the
parameter $\xi g^2$ for which two-loop/one-loop corrections are 
small (and perturbation theory valid), 
the uncertainty in the  bound mass from the gauge
dependence of the result is small and expected to be, for the case of
the Standard Model, inside the range of (\ref{error}).}
in the lower bound. All together we have estimated the total uncertainty as
\be
\label{error}
\Delta M_H \simlt 5\ {\rm GeV}
\ee

%
\subsection{Metastability bounds}
 
In the last subsection we have seen that 
in the region of Fig.~\ref{fval2}
below the dashed line the standard EW minimum is
metastable. However we should not draw physical consequences 
from this fact since we still do not
know at which minimum does the Higgs field sit. Thus, the real physical 
constraint we have to impose is avoiding
the Higgs field sitting at its non-standard minimum. 
In fact the Higgs field can be sitting at its 
zero temperature non-standard minimum if:
\begin{enumerate}
\item
The Higgs field was driven from the origin to the non-standard minimum 
at finite temperature 
by thermal fluctuations in a non-standard EW phase transition at 
high temperature. 
This minimum evolves naturally to the non-standard minimum at zero 
temperature. In this case 
the standard EW phase transition, at $T\sim 10^2$ GeV, will not take place.
\item
The Higgs field was driven from the origin to the 
standard minimum at $T\sim 10^2$ GeV, but decays,
at zero temperature, to the non-standard minimum by a quantum fluctuation.
\end{enumerate}
In Fig.~\ref{fval1} we have depicted the 
effective potential at $T=2.5\times 
10^{15}$ GeV (thin solid line) which
is the corresponding 
transition temperature. Our finite temperature 
potential~\cite{EQtemp} incorporates plasma effects~\cite{Q} 
by one-loop resummation of Debye masses~\cite{DJW}.  The tunnelling 
probability per unit time per
unit volume  was computed long ago for thermal~\cite{Linde} and 
quantum~\cite{Coleman} fluctuations.
At finite temperature it is given by $\Gamma/\nu\sim T^4 \exp(-S_3/T)$, 
where $S_3$ is the euclidean action evaluated
at the bounce solution $\phi_B$. The semiclassical 
picture is that unstable bubbles are nucleated behind the
barrier at $\phi_B(0)$ with a probability given by $\Gamma/\nu$. Whether 
or not they fill the Universe depends on
the relation between the probability rate and the expansion 
rate of the Universe. By normalizing the former
with respect to the latter we obtain a normalized probability $P$, 
and the condition for decay corresponds
to $P\sim 1$. Of course our results are trustable,  
and the decay actually happens, only if
$\phi_B(0)<\Lambda$, so that the similar condition to (\ref{condstab}) is 
\be
\label{condmeta}
\Lambda< \phi_B(0)
\ee
The condition of no-decay (metastability condition) has 
been plotted in Fig.~\ref{fval2}  (solid lines)
for $\Lambda=10^{19}$ GeV and 10 TeV. The region
between the dashed and the solid line corresponds to 
a situation where the non-standard minimum exists
but there is no decay to it at finite temperature. 
In the region below the solid lines the Higgs field is sitting
already at the non-standard minimum at $T\sim 10^2$ GeV, and the  standard EW 
phase transition does not happen.
A fit for $\Lambda=10^{19}$ GeV is
\be
\label{cota19meta}
M_H[GeV] > 125+2.28 (M_t[GeV]-175)-4.89 \frac{\alpha_s(M_Z)-0.12}
{0.006}
\ee

As for the gauge dependence, we expect it to affect very little the
previous results. The total potential at finite temperature is entirely
dominated, for $\phi<\phi_B(0)$, by the thermal correction 
(see Fig.~\ref{fval1}). But the one-loop 
thermal correction coming from gauge boson and the 
top-quark propagation, 
as well as the improvement due to
thermal masses, is gauge independent~\cite{Kapusta}, while all the
gauge dependence is encoded in the contribution from Higgs and Goldstone
bosons, which would affect very little the result.

We also have evaluated the tunnelling probability 
at zero temperature from the standard EW minimum to the
non-standard one. The result of the calculation 
should translate, as in the previous case, in lower bounds
on the Higgs mass for differentes 
values of $\Lambda$. The corresponding bounds are shown 
in Fig.~\ref{fval2} in
dotted lines. Since the dotted lines 
lie always below the solid ones, the possibility of quantum tunnelling at
zero temperature does not impose any extra constraint.

As a consequence of all improvements in the 
calculation, our bounds are lower than previous 
estimates~\cite{AV}. To fix ideas, for $M_t=175$ GeV, the 
bound reduces by $\sim 10 $ GeV for $\Lambda=10^4$ GeV,
and $\sim 30$ GeV for $\Lambda=10^{19}$ GeV.

\subsection{Perturbativity bounds}

Up to here we have described lower bounds on the Higgs mass
based on stability arguments. Another kind of bounds, which have
been used in the literature, are upper bounds based on the
requirement of perturbativity of the SM up to the high scale
(the scale of new physics) $\Lambda$. 

\begin{figure}[hbt]
\centerline{
\psfig{figure=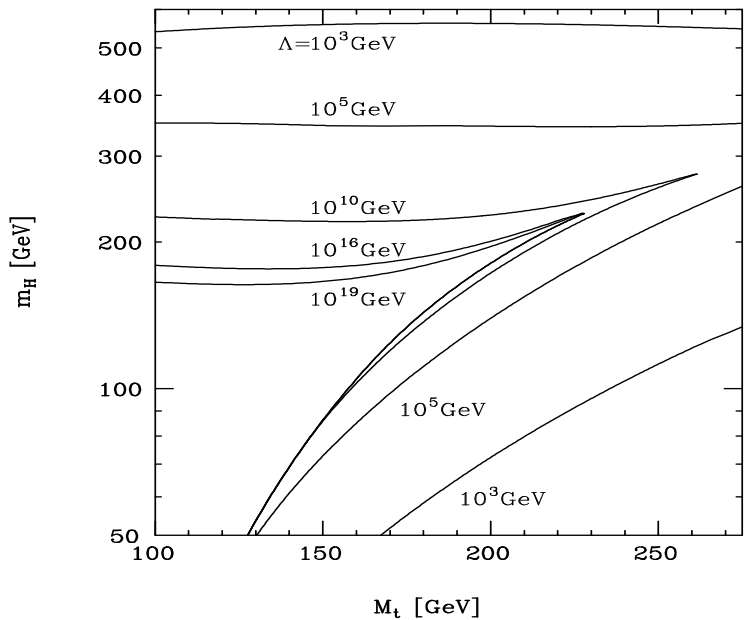,height=10cm,width=17cm,angle=90}}
\caption{Perturbativity and stability bounds on the SM Higgs
boson. $\Lambda$ denotes the energy scale where the particles
become strongly interacting.}
\label{lepp}
\end{figure}

\begin{figure}[hbt]
\centerline{
\psfig{figure=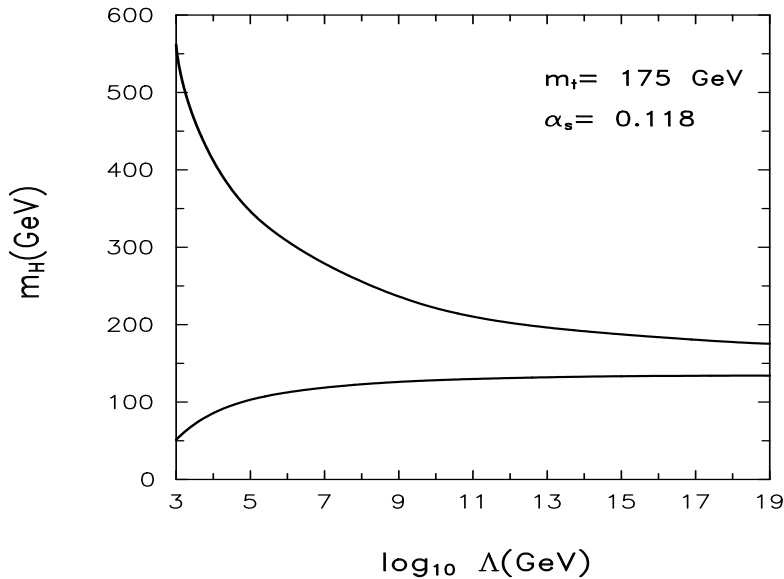,height=7.5cm,width=7cm,bbllx=7.cm,bblly=4.cm,bburx=16.5cm,bbury=17cm}}
\caption{Perturbativity and stability bounds on the SM Higgs
boson mass as a function of $\Lambda$, for $M_t=175$ GeV.}
\label{marce2}
\end{figure}

Since the quartic coupling grows with the scale~\footnote{In fact
the value of the renormalization scale where the quartic
coupling starts growing depends on the value of the top-quark
mass.}, it will blow up to infinity at a given scale: the scale where
$\lambda$ has a Landau pole. The position of the Landau pole
$\Lambda$ is, by definition, the maximum scale up to which the
SM is perturbatively valid. In this way assuming the SM remains
valid up to a given scale $\Lambda$ amounts to requiring an
upper bound on the Higgs mass from the perturbativity 
condition~\cite{LEP2} 
\be
\label{perturbcond}
\frac{\lambda(\Lambda)}{4\pi}\leq 1
\ee
This upper bound depends on the scale $\Lambda$ and very mildly
on the top-quark mass $M_t$ through its influence on the
renormalization group equations of $\lambda$. We have plotted in
Fig.~\ref{lepp} this upper bound for different values of the high scale
$\Lambda$, along with the corresponding stability bounds, and in
Fig.~\ref{marce2} both bounds as functions of $\Lambda$ for $M_t=175$
GeV. 

\section{The Minimal Supersymmetric Standard Model}

The Minimal Supersymmetric Standard Model 
(MSSM)~\cite{susy} is the best
motivated extension of the SM where some of their theoretical problems 
(e.g. the hierarchy problem inherent with the fact that the
SM cannot be considered as a fundamental theory for energies
beyond the Planck scale) find (at least) a technical solution. 
The Higgs sector of the MSSM~\cite{Hunter} requires two Higgs doublets, with
opposite hypercharges, as
\begin{eqnarray}
\label{higgsmssm}
H_1 & = & \left(
\begin{array}{c}
H_1^0 \\
H_1^-
\end{array}
\right)_{-1/2} \nonumber \\
H_2 & = & \left(
\begin{array}{c}
H_2^+ \\
H_2^0
\end{array}
\right)_{1/2}
\end{eqnarray}
The reason for this duplicity is twofold. On the one hand it is
necessary to cancel the triangular anomalies generated by the
higgsinos. On the other hand it is required by the structure of
the supersymmetric theory to give masses to all fermions.

The most general gauge invariant scalar potential is given,
for a general two-Higgs doublet model, by:
\begin{eqnarray}
\label{higgs2}
V& = & m_1^2 |H_1|^2+m_2^2|H_2|^2+(m_3^2 H_1 H_2+h.c.)
+\frac{1}{2}\lambda_1(H_1^\dagger H_1)^2\nonumber\\
&&+\frac{1}{2}\lambda_2(H_2^\dagger
H_2)^2+\lambda_3(H_1^\dagger H_1)(H_2^\dagger H_2)
+\lambda_4(H_1 H_2)(H_1^\dagger H_2^\dagger) \\
&&+\left\{\frac{1}{2}\lambda_5(H_1 H_2)^2+
\left[\lambda_6(H_1^\dagger H_1)+\lambda_7(H_1^\dagger
H_2^\dagger) \right](H_1 H_2)+h.c.\right\} \nonumber
\end{eqnarray}
However, supersymmetry provides the following tree-level
relations between the previous couplings:
\begin{eqnarray}
\label{lambdastree}
&&\lambda_1  = \lambda_2=\frac{1}{4}(g^2+g'^2) \nonumber\\
&&\lambda_3  = \frac{1}{4}(g^2-g'^2),\ 
\lambda_4  = -\frac{1}{4}g^2  \nonumber\\
&&\lambda_5  =  \lambda_6= \lambda_7=0 
\end{eqnarray}
Replacing (\ref{lambdastree}) into (\ref{higgs2}) one obtains
the tree-level potential of the MSSM, as:
\begin{eqnarray}
\label{potmssm}
V_{\rm MSSM}& = & m_1^2 H_1^\dagger H_1+m_2^2 H_2^\dagger H_2
+m_3^2(H_1 H_2+h.c.) \\
&&+\frac{1}{8}g^2\left(H_2^\dagger \vec{\sigma} H_2+
H_1^\dagger\vec{\sigma}H_1\right)^2
+\frac{1}{8}g'^2\left(H_2^\dagger H_2-H_1^\dagger H_1\right)^2
\nonumber
\end{eqnarray}

This potential, along with the gauge and Yukawa couplings in the
superpotential,
\be
\label{superpotential}
W=h_u Q\cdot H_2 U^c+h_d Q\cdot H_1 D^c+ h_\ell L\cdot H_1 E^c
+\mu H_1\cdot H_2
\ee
determine all couplings and masses (at the tree-level) of the
Higgs sector in the MSSM.

\subsection{Tree-level mass relations}

After gauge symmetry breaking,
\begin{eqnarray}
v_1 & = & \langle {\rm Re}\; H_1^0 \rangle \nonumber \\
v_2 & = & \langle {\rm Re}\; H_2^0 \rangle 
\end{eqnarray}
the Higgs spectrum contains one neutral CP-odd Higgs $A$
(with mass $m_A$, that will be taken as a free parameter)
\be
A=\cos\beta\;{\rm Im}H_2^0+\sin\beta\;{\rm Im}H_1^0
\ee
and one neutral Goldstone $\chi^0$
\be
\chi^0=-\sin\beta\;{\rm Im}H_2^0+\cos\beta\;{\rm Im}H_1^0
\ee
with $\tan\beta=v_2/v_1$. It also contains one complex charged
Higgs $H^\pm$,
\be
H^+=\cos\beta\; H_2^+ +\sin\beta\;(H_1^-)^*
\ee
with a (tree-level) mass
\begin{eqnarray}
 m^2_{H^{\pm}} &=& m_A^2 + (\lambda_5 - \lambda_4) v^2,
\label{mhch}
\end{eqnarray}   
and one charged Goldstone $\chi^\pm$,
\be
\chi^+=-\sin\beta\; H_2^+ +\cos\beta\;(H_1^-)^*.
\ee
Finally the Higgs spectrum contains two CP-even neutral Higgs
bosons $H,{\cal H}$ (the light and the heavy mass eigenstates)
which are linear combinations of Re~$H_1^0$ and
Re~$H_2^0$, with a mixing angle $\alpha$ given by
\begin{eqnarray}
      \sin 2\alpha = \frac{2M_{12}^2}
{\sqrt{\left(Tr M^2\right)^2-4\det M^2}}
 \end{eqnarray}
\begin{eqnarray}
      \cos 2\alpha = \frac{M_{11}^2-M_{22}^2}
{\sqrt{\left(Tr M^2\right)^2-4\det M^2}}
\end{eqnarray}     
and masses

\begin{eqnarray}
m^2_{H({\cal H})} &=& \frac{Tr M^2 \mp \sqrt{(TrM^2)^2 - 4 \det M^2}}{2}
\label{mhH}
\end{eqnarray}   
where
\begin{equation}  
         TrM^2 =  M_{11}^2 + M_{22}^2 \;\; ; \;\;\;\;\;
     \det M^2 = M_{11}^2 M_{22}^2 - \left( M_{12}^2 \right)^2,
\label{detm2}
\end{equation}
with
 \begin{eqnarray}
  M^2_{12} &=&  2 v^2 [\sin \beta \cos \beta (\lambda_3 + \lambda_4) + 
     \lambda_6 \cos^2 \beta + \lambda_7 \sin^2 \beta ] -
     m_A^2 \sin \beta \cos \beta
 \nonumber\\    
     M^2_{11} &=& 2 v^2 [\lambda_1  \cos^2 \beta + 2 
     \lambda_6  \cos \beta \sin \beta 
     + \lambda_5 \sin^2 \beta] + m_A^2 \sin^2 \beta \\
     M^2_{22} &=&  2 v^2 [\lambda_2 \sin^2 \beta +2 \lambda_7  \cos \beta
     \sin \beta + \lambda_5 \cos^2 \beta] + m_A^2 \cos^2 \beta . \nonumber
 \end{eqnarray}  
where ($v^2=v_1^2+v_2^2$) the couplings $\lambda_i$ are taken 
at their tree level values
(\ref{lambdastree}), yielding
\be
\label{masapm}
m_{H^\pm}^2=M_W^2+m_A^2
\ee
and
\be
\label{masahH}
m^2_{H,{\cal H}}=\frac{1}{2}\left[
m_A^2+M_Z^2\mp\sqrt{(m_A^2+M_Z^2)^2-4m_A^2M_Z^2\cos^2 2\beta} .
\right]
\ee

\subsection{The Higgs tree level couplings}

All couplings in the Higgs sector are functions of the gauge
($G_F,g,g'$) and Yukawa couplings, as in the SM, and of the
previously defined mixing angles $\beta,\alpha$. 
Some relevant couplings are contained in Table~1
where all particle momenta, in squared brackets, are incoming,
and $\varphi\equiv(H,{\cal H},A)$.

\vspace{1cm}
\begin{center}
\begin{tabular}{|c|c|}\hline
Vertex & Couplings \\ \hline
& \\
$(H,{\cal H})WW $ & $(\phi WW)_{\rm
SM}[\sin(\beta-\alpha),\cos(\beta-\alpha)]$ \\
& \\
$(H,{\cal H})ZZ $ & $(\phi ZZ)_{\rm
SM}[\sin(\beta-\alpha),\cos(\beta-\alpha)]$ \\
& \\
$\varphi[p]W^\pm H^\mp [k] $ & $\mp i\frac{g}{2}(p+k)^\mu
[\cos(\beta-\alpha), -\sin(\beta-\alpha),\pm i]$ \\
& \\
$\varphi u\bar{u} $  & $(\phi u\bar{u})_{\rm
SM}[{\displaystyle \frac{\cos\alpha}{\sin\beta},
\frac{\sin\alpha}{\sin\beta} , -i\gamma_5 \cot\beta]} $  \\
& \\
$\varphi d\bar{d} $  &$(\phi d\bar{d})_{\rm
SM}[{\displaystyle -\frac{\sin\alpha}{\cos\beta},
\frac{\cos\alpha}{\cos\beta}, -i\gamma_5 \tan\beta] } $ \\
& \\
$H^- u\bar{d} $
& $ {\displaystyle
\frac{ig[(m_d\tan\beta+m_u\cot\beta) - (m_d
\tan\beta -m_u \cot\beta)\gamma_5] }{2\sqrt{2}M_W}} $ \\
& \\
$ H^+ \bar{u} d $ & $ {\displaystyle 
\frac{ig[(m_d\tan\beta+m_u\cot\beta) + (m_d
\tan\beta -m_u \cot\beta)\gamma_5] }{2\sqrt{2}M_W}} $ \\
& \\
$(\gamma,Z)H^+[p]H^- [k] $ & $ {\displaystyle 
-i(p+k)^\mu\left[e,g\frac{\cos
2\theta_W}{2 \cos\theta_W}\right] } $ \\
& \\
$h[p] A [k] Z $ & ${\displaystyle 
-\frac{e}{2\cos\theta_W\sin\theta_W} (p+k)^\mu
\cos(\beta-\alpha) } $ \\
& \\ \hline
\end{tabular}
\end{center}

\vspace{0.5cm}
\begin{center}
Table 1
\end{center}

\subsection{Radiatively corrected masses}

The mass spectrum satisfies the following
tree-level relations: 
\begin{eqnarray}
\label{treerel}
m_H & < & M_Z|\cos 2\beta| \nonumber \\
m_H & < & m_A \\
m_{H^\pm} & > & M_W \nonumber
\end{eqnarray}
which could have a number of very important phenomenological
implications, as it is rather obvious. However, it was
discovered~\cite{Effpot}$^-$\cite{Andrea} 
that radiative corrections are important and can
spoil the above tree level relations with a great
phenomenological relevance. A detailed knowledge of radiatively
corrected couplings and masses is necessary for experimental
searches in the MSSM.

The {\bf effective potential} methods to compute the (radiatively
corrected) Higgs mass spectrum in the 
MSSM are useful since they allow to {\bf resum}
(using Renormalization Group (RG) techniques) LL,
NTLL,..., corrections to {\bf all orders}
in perturbation theory. These methods~\cite{Effpot,EQ}, 
were first developed in the early nineties.

Effective potential methods are based on the {\bf run-and-match}
procedure by which all dimensionful and dimensionless couplings
are running with the RGE scale, for scales greater than the
masses involved in the theory. When the RGE scale
equals a particular mass threshold, heavy fields decouple,
eventually leaving threshold effects in order to match the
effective theory below and above the mass threshold. For
instance, assuming a common soft supersymmetry breaking mass 
for left-handed and right-handed stops and sbottoms, 
$M_S\sim m_Q\sim m_U\sim m_D$, and assuming for the top-quark mass, 
$m_t$, and for the CP-odd Higgs mass, $m_A$, the range 
$m_t\leq m_A\leq M_S$, we have: for scales $Q\geq M_S$, the MSSM, for
$m_A\leq Q\leq M_S$ the two-Higgs doublet model (2HDM), and for
$m_t\leq Q\leq m_A$ the SM. Of course there are
thresholds effects at $Q=M_S$ to match the MSSM with the 2HDM, and
at $Q=m_A$ to match the 2HDM with the SM. 

\subsubsection{The case $m_A\sim M_S$}

The case $m_A\sim M_S$ is, not only a great simplification since the effective
theory below $M_S$ is the SM, but also of great interest, since it
provides the upper bound on the mass of the lightest Higgs
(which is interesting for phenomenological purposes, e.g. at
LEP2). In this case we have found~\cite{CEQW,HHH} that, 
in the absence of mixing
(the case $X_t=0$) two-loop corrections resum in the one-loop
result shifting the energy scale from $M_S$ (the tree-level scale)
to $\sqrt{M_S\; m_t}$. More explicitly,
\be
\label{resum}
m_H^2=M_Z^2 \cos^2 2\beta\left(1-\frac{3}{8\pi^2}h_t^2\; t\right)
+\frac{3}{2\pi^2 v^2}m_t^4(\sqrt{M_S m_t}) t
\ee
where $t=\log(M_{\rm S}^2/M_t^2)$ and  bottom Yukawa coupling is neglected.

\begin{figure}[htb]
\centerline{
\psfig{figure=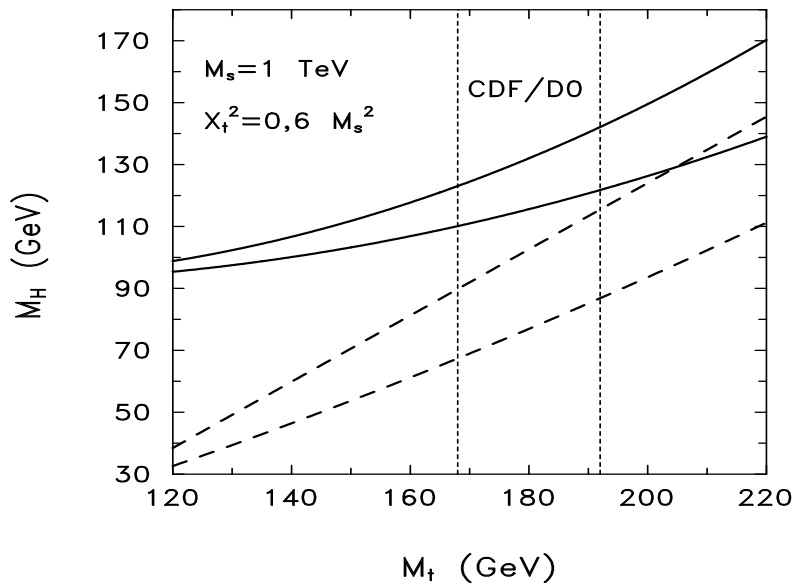,height=7.5cm,width=7cm,bbllx=5.5cm,bblly=3.cm,bburx=15.cm,bbury=16.cm}}
\caption{Plot of $M_H$ as a function of $M_t$ for $\tan\beta\gg 1$
(solid lines), $\tan\beta=1$ (dashed lines), and $X_t^2=6 M_S^2$ (upper set),
$X_t=0$ (lower set).}
\label{fval3}
\end{figure}

In the presence of mixing ($X_t\neq 0$), the run-and-match procedure
yields an extra piece in the SM effective potential 
$\Delta V_{\rm th}[\phi(M_S)]$ whose second derivative gives an
extra contribution to the Higgs mass, as
\be
\label{Deltathm}
\Delta_{\rm th}m_H^2=\frac{\partial^2}{\partial\phi^2(t)}
\Delta V_{\rm th}[\phi(M_S)]=
\frac{1}{\xi^2(t)}
\frac{\partial^2}{\partial\phi^2(t)}
\Delta V_{\rm th}[\phi(M_S)]
\ee
which, in our case, reduces to 
\be
\label{masthreshold}
\Delta_{\rm th}m_H^2=
\frac{3}{4\pi^2}\frac{m_t^4(M_S)}{v^2(m_t)}
\frac{X_t^2}{M_S^2}\left(2-\frac{1}{6}\frac{X_t^2}{M_S^2}\right)
\ee
where $X_t=(A_t-\mu/\tan\beta)$ is the mixing in the stop mass
matrix, the parameters $A_t$ and $\mu$ being the trilinear
soft-breaking coupling in the stop sector and the supersymmetric
Higgs mixing mass, respectively. The maximum of
(\ref{masthreshold}) corresponds to $X_t^2=6 M_S^2$ which provides
the maximum value of the lightest Higgs mass: this case will be
referred to as the case of maximal mixing.

We have plotted in Fig.~\ref{fval3} the lightest Higgs pole mass 
$M_H$, where all NTLL corrections
are resummed to all-loop by the RGE, 
as a function of $M_t$~\cite{CEQR}. From Fig.~\ref{fval3} we
can see that the present experimental band from CDF/D0 for the
top-quark mass requires $M_H\simlt 135$ GeV, while if we fix
$M_t=175$ GeV, the upper bound $M_H\simlt 125$ GeV
follows. It goes without saying
that these figures are extremely relevant for MSSM Higgs searches 
at LEP2.

\subsubsection{The case $m_A\simlt M_S$}

The case $m_A\simlt M_S$ is more complicated since the effective theory
below the supersymmetric scale $M_S$ is the 2HDM
with $\lambda_i$ couplings in (\ref{higgs2}). However since radiative
corrections in the 2HDM are equally dominated by the top-quark, we can
compute analytical expressions based upon the two loop LL approximation
at the scale $Q^2\sim M_t^2$. 
Our approximation~\cite{CEQW} differs from
the LL all-loop numerical resummation by $\simlt 3$ GeV, which we
consider the uncertainty inherent in the theoretical calculation,
provided the mixing is moderate and, in particular, bounded by the
condition,
\be
\label{condicion}
\left|\frac{m^2_{\;\widetilde{t}_1}-m^2_{\;\widetilde{t}_2}}
{m^2_{\;\widetilde{t}_1}+m^2_{\;\widetilde{t}_2}}\right|\simlt 0.5
\ee
where $\widetilde{t}_{1,2}$ are the two stop mass eigenstates.

The above quartic couplings are given by~\footnote{For 
simplicity, we neglect the leading $D$-term contributions
and bottom Yukawa coupling effects. The latter may become large for values of 
$\tan\beta \simeq m_t/m_b$, where $m_b$ is the running
bottom mass at the scale $M_t$. Complete formulae have been
worked out~\cite{CEQW}. }
   
\begin{eqnarray}
\Delta  \lambda_1 & =& 
-    \frac{3}{96\pi^2} \; h_t^4\;\frac{\mu^4}{M_{\rm S}^4}
        \left[ 1+ \frac{1}{16 \pi^2} \left( 9\;h_t^2 
     -  16 g_s^2 \right) t  \right] \\
\label{lambda}  
 \Delta\lambda_2& =& \frac{3}{8 \pi^2}\; h_t^4\; \left[
         t + \frac{X_{t}^0}{2} + \frac{1}{16 \pi^2}
        \left( \frac{3 \;h_t^2}{2}        
     - 8\; g_s^2 \right) \left( X_{t}^0\;t + t^2\right) \right] \\ 
\Delta\lambda_3 &=& \frac{3}{96\pi^2} \; 
h_t^4\; \left[\frac{3 \mu^2}{M_{\rm S}^2}
                - \frac{\mu^2 A_t^2}{M_{\rm S}^4} \right]
        \left[ 1+ \frac{1}{16 \pi^2} \left (6\;h_t^2 
     -  16 g_s^2 \right) t  \right]\\
\Delta\lambda_4 &=&
\frac{3}{96\pi^2} \; h_t^4\; \left[\frac{3 \mu^2}{M_{\rm S}^2}
                         - \frac{\mu^2 A_t^2}{M_{\rm S}^4} \right]
        \left[ 1+ \frac{1}{16 \pi^2} \left (6\;h_t^2 
     -  16 g_s^2 \right) t  \right]\\
\Delta \lambda_5 &=& -\; \frac{3}{96\pi^2} \; h_t^4\; 
                   \frac{\mu^2 A_t^2}{M_{\rm S}^4} 
        \left[ 1+ \frac{1}{16 \pi^2} \left ( 6\;h_t^2 
     -  16 g_s^2 \right) t  \right]   \\
\Delta\lambda_6 &=& \frac{3}{96 \pi^2}\; h_t^4\;
       \frac{\mu^3 A_t}{M_{\rm S}^4}
   \left[1+ \frac{1}{16\pi^2}
   \left(\frac{15}{2} h_t^2 - 16 g_s^2 \right)
        t \right]  \\
\Delta\lambda_7 &=  &   
\frac{3}{96 \pi^2}\; h_t^4\;\frac{\mu}{M_{\rm S}} 
\left(\frac{A_t^3}{M_{\rm S}^3}
         - \frac{6 A_t}{M_{\rm S}}\right)
   \left[1+ \frac{1}{16\pi^2}
\left(\frac{9}{2} h_t^2 - 16 g_s^2 \right)
       t  \right],  
\end{eqnarray}
%
\begin{figure}[ht]
\centerline{
\psfig{figure=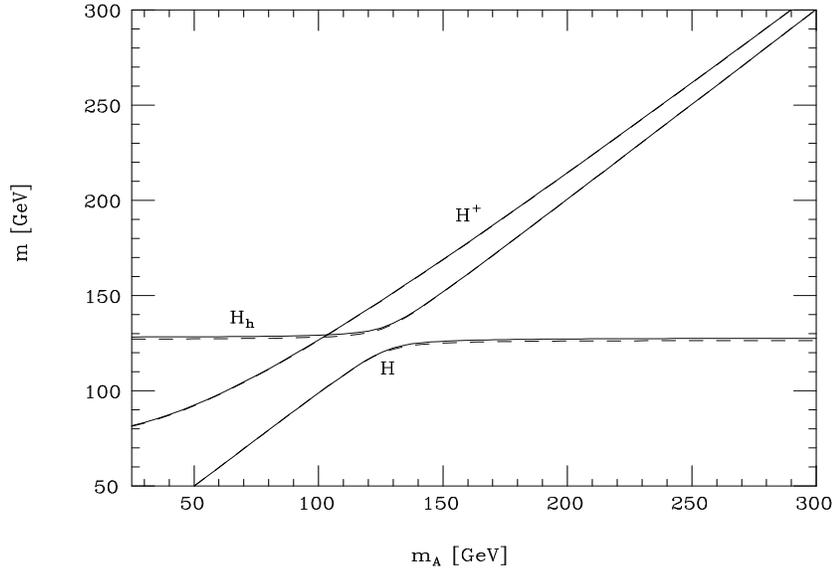,height=9.5cm,width=14cm,angle=90} }
\caption[0]
{The neutral ($H,{\cal H}\equiv H_h$ in the figure) 
and charged ($H^+$) Higgs mass spectrum 
as a function of the CP-odd Higgs mass $m_A$ for
a physical top-quark mass $M_t =$ 175 GeV and $M_S$ = 1 TeV, as
obtained from the one-loop improved RGE evolution 
(solid lines) and the analytical formulae (dashed lines).
All sets of curves correspond to
$\tan \beta=$ 15 and large squark mixing, $X_t^2 = 6 M_S^2$
($\mu=0$).} 
\label{fval4}
\end{figure}

\noindent 
where we have defined,
\begin{eqnarray}
X_{t}^0 & = & \frac{2 A_t^2}{M_{\rm S}^2}
                  \left(1 - \frac{A_t^2}{12 M_{\rm S}^2} \right).
\end{eqnarray}

All quantities in the approximate formulae are defined at the scale
$M_t$. For $m_A\leq M_t$,  $\tan\beta$ is fixed at the scale $m_A$,
while for $m_A\geq M_t$,  $\tan\beta$ is given by~\cite{Andrea}
\be
\label{tanbeta}
\tan\beta(M_t)=\tan\beta(m_A)\left[1+\frac{3}{32\pi^2} h_t^2 
\log\frac{m_A^2}{M_t^2}\right].
\ee
For the case in which the CP-odd Higgs mass $m_A$ is lower than
$M_{\rm S}$, but still  larger than the top-quark mass scale,
we decouple, in the numerical computations, the heavy Higgs doublet
and define an effective quartic coupling for the light Higgs,
which is related to the running  Higgs mass at the scale $m_A$
through
\begin{equation}
\lambda(m_A) = \frac{m_h(m_A)}{2 v^2}.
\end{equation}
The low energy value of the quartic coupling is then obtained by
running the SM renormalization-group equations from the scale
$m_A$ down to the scale $M_t$. We have defined up to here the
running masses. To obtain physical masses we have to use
(\ref{masaspolo2}) with $\Delta\Pi$ defined in appendix B.

Notice that the knowledge of the radiatively corrected quartic couplings
$\lambda_i$, $i=1,\dots, 7$, and hence of 
the corresponding value of the Higgs mixing
angle $\alpha$, permits the  evaluation of all 
radiatively corrected Higgs couplings in Table~1. 
In Fig.~\ref{fval4} the Higgs mass spectrum is plotted versus $m_A$.

There are two possible caveats in the approximation we have just 
presented: {\bf i)} Our expansion parameter $\log(M_S^2/m_t^2)$
does not behave properly in the supersymmetric limit $M_S\rightarrow 0$,
where we should recover the tree-level result. {\bf ii)} We have expanded
the threshold function $\Delta V_{\rm th}[\phi(M_S)]$ to order $X_t^4$.
In fact keeping the whole threshold function $\Delta V_{\rm th}[\phi(M_S)]$
we would be able to go to larger values of $X_t$ and to evaluate the
accuracy of the approximation (\ref{masthreshold}).
Only then we will be able to
check the reliability of the maximum value of the
lightest Higgs mass (which corresponds to the maximal mixing) as provided
in the previous sections.
This procedure has been properly followed in Refs.~\cite{CEQW} 
and~\cite{CQW}, where the most general case $m_Q\neq m_U\neq m_D$ has been
considered. We have proved that keeping the exact threshold function
$\Delta V_{\rm th}[\phi(M_S)]$, and properly running its value from the
high scale to $m_t$ with the corresponding anomalous dimensions as in
(\ref{Deltathm}), produces two effects: {\bf i)} It makes a resummation
from $M_S^2$ to $M_S^2+m_t^2$ and generates as (physical) expansion
parameter $\log[(M_S^2+m_t^2)/m_t^2]$. {\bf ii)} It generates a whole
threshold function $X_t^{\rm eff}$ such that (\ref{masthreshold})
becomes
\be
\label{masthreshold2}
\Delta_{\rm th}m_H^2=
\frac{3}{4\pi^2}\frac{m_t^4[M_S^2+m_t^2]}{v^2(m_t)}
X_t^{\rm eff}
\ee
and 
\be
\label{desarrollo}
X_t^{\rm eff}=\frac{X_t^2}{M_S^2+m_t^2}
\left(2-\frac{1}{6}\frac{X_t^2}{M_S^2+m_t^2}\right)+\cdots
\ee
%
\begin{figure}[htb]
\centerline{
\psfig{figure=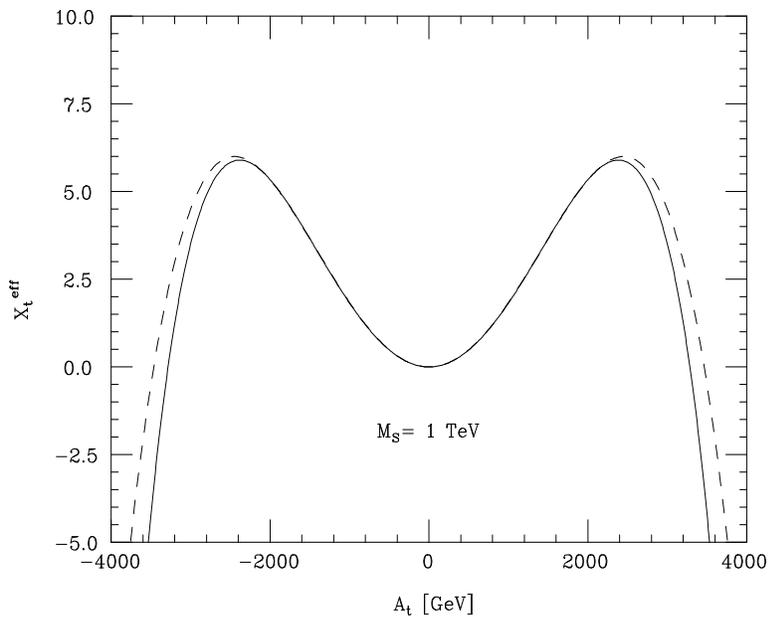,width=13cm,height=10.5cm,angle=90}}
\caption[0]{Plot of the exact (solid line) and approximated
(dashed line) effective mixing $X_t^{\rm eff}$ as a function of $A_t$,
for $M_S=1$ TeV and $\mu=0$.}
\label{estafigura}
\end{figure} 
%
In fact we have plotted $X_t^{\rm eff}$ as a function of $A_t$ (solid line)
and compared with the approximation where we keep only terms up to
$X_t^4$ (dashed line), as we did in the previous sections. 
The result shows that the
maximum of both curves are very close to each other, what justifies
the reliability of previous upper bounds on the lightest Higgs mass
as, e.g., in Fig.~\ref{estafigura}

\section{Non-minimal Supersymmetric Standard Models}

The tree-level bound (\ref{treerel}) 
does no longer hold in supersymmetric models
with extra Higgs fields, the so-called NMSSM. We will study in this
section Higgs bounds in a general class of NMSSM.

The first (obvious) enlargement of the Higgs sector consists in adding
pairs of Higgs doublets $H_1^{(j)}$, $H_2^{(j)}$, $j=1,...,N$. These
models have been analyzed~\cite{6}
and their lightest scalar Higgs boson shown to
have the tree-level bound (\ref{treerel}).

Consider now NMSSM with Higgs doublets
$H_1$, $H_2$ and neutral scalar fields
$N_{12}^{(i)},\ N_{11}^{(j)}$,
$\ N_{22}^{(j)}$ (either $SU(2)_L
\times U(1)_Y$ singlets or making part of higher dimensional
representations) with a cubic superpotential $f=g+f_{YUK}$
\begin{equation}
g= \vec{\lambda}\cdot\vec{N}_{12}H_1^o H_2^o +
\sum_{i=1}^2 \vec{\chi}_{i}\cdot\vec{N}_{ii}(H_i^o)^2 ,
\label{superp}
\end{equation}
where $\vec{\lambda}\cdot\vec{N}\equiv
\sum_j \lambda_j N^{(j)}$ and $f_{YUK}$ contains all Yukawa
couplings giving mass to fermions.
Then, the lightest scalar Higgs boson mass has an upper bound given 
by~\cite{7,KKW}
\begin{equation}
\displaystyle{\frac{m_h^2}{v^2}}\leq \frac{1}{2}(g^2+g'^2)\cos^22\beta
+\vec{\lambda}^2\sin^22\beta
+\vec{\chi}_{1}^2\cos^4\beta +
\vec{\chi}_{2}^2\sin^4\beta.
\label{genbound}
\end{equation}

The bound for the MSSM is recovered from (\ref{genbound}) when
$\vec{\lambda}=\vec{\chi}_{1}=\vec{\chi}_{2}=0$.
However in NMSSM some of the Yukawa couplings
in (\ref{superp}) can be non-zero. In that case the upper bound on the
lightest scalar Higgs boson mass comes from the requirement that the
supersymmetric theory remains perturbative up to some scale
$\Lambda$, in the energy range where the theory holds.

We will keep in $f_{YUK}$ the top and bottom quark Yukawa couplings, $i. e.$
\begin{equation}
f_{YUK}=h_t  Q\cdot H_2 U^c +h_b  Q\cdot H_1 D^c,
\label{superY}
\end{equation}
with boundary conditions
\begin{equation}
\begin{array}{cc}
h_t={\displaystyle\frac{g}{\sqrt{2}}\frac{m_t}{m_W}}(1+\cot^2\beta)^{1/2},&
h_b={\displaystyle\frac{g}{\sqrt{2}}\frac{m_b}{m_W}}(1+\tan^2\beta)^{1/2}.
\end{array}
\label{Yukawas}
\end{equation}
$m_t$ in (\ref{Yukawas}) 
will be considered as a variable while $h_b$ is fixed by
$m_b$, which is taken to be $m_b(2\ m_b)=\ 5\ GeV$. For $\tan
\beta\gg 1$, $h_b$ can become important. In particular it is
comparable to $h_t$ for $\tan\beta\sim m_t/m_b$. $h_{\tau}$ will be
neglected since it is given by $h_b (m_{\tau}/m_b)$ for all values
of $\tan\beta$. The cubic $g$-superpotential in (\ref{superp}) and so the
tree-level mass in (\ref{genbound}) are model dependent. 
The latter depends on the
couplings $\vec{\lambda}$, $\vec{\chi}_i$
allowed by the perturbative requirement.

\subsection{NMSSM with an arbitrary number of singlets}

These models are particularly interesting in the sense that the
prediction of perturbative unification of the MSSM from LEP precision
measurements is not spoiled by the extra matter.
They are defined by a Higgs sector containing $H_1$, $H_2$
and $n$ singlets $S_i\ (i=1,...,n)$ with a cubic superpotential
\begin{equation}
g= \vec{\lambda}\cdot \vec{S} H_1\cdot H_2 + \frac{1}{6} \sum_{i,j,k}
\chi_{ijk} S_iS_jS_k.
\end{equation}
The model with $n=1$ has been studied in great detail in the
literature~\cite{6}$^-$\cite{singlet3}. 
The tree-level upper bound on the mass of the
lightest scalar Higgs boson for the case of arbitrary $n$ can be written
as~\cite{7,10}:
\begin{equation}
m_h^2\leq {\displaystyle\left(\right.} \cos^2 2\beta
+ {\displaystyle\frac{2\vec{\lambda}^2
\cos^2\theta_W}{g^2}}\sin^22\beta \left.\right)m_Z^2.
\end{equation}
 The relevant one-loop RGE are

\be
\begin{array}{c}
 4\pi^2\dot{\vec{\lambda}^2}=\left\{ -\frac{3}{2}g^2 - \frac{1}{2}g'^2
 +2\vec{\lambda}^2  + \frac{3}{2}(h_t^2 + h_b^2)\right\}\vec{\lambda}^2
 + \frac{1}{4}\lambda_i\lambda_j tr(M_iM_j),\vspace{.4cm}\\
 8\pi^2\dot M_k=3\lambda_k \vec{M}\cdot\vec{\lambda} + \frac{3}{4}
 tr(\vec{M}\cdot M_k)\cdot \vec{M},\vspace{.4cm}\\
 8\pi^2\dot h_t=\left\{ -\frac{3}{2}g^2 - \frac{13}{18}g'^2
 - \frac{8}{3}g_s^2
   +\frac{1}{2}\vec{\lambda}^2+ 3h_t^2 + \frac{1}{2} h_b^2\right\}h_t,
 \vspace{.4cm}\\
 8\pi^2\dot h_b=\left\{ -\frac{3}{2}g^2 - \frac{7}{18}g'^2
 - \frac{8}{3}g_s^2
   +\frac{1}{2}\vec{\lambda}^2+ \frac{1}{2}h_t^2
 + 3 h_b^2\right\}h_b,\vspace{.4cm}\\
\end{array}
\label{RGE}
\ee

\begin{figure}[htb]
\centerline{
\psfig{figure=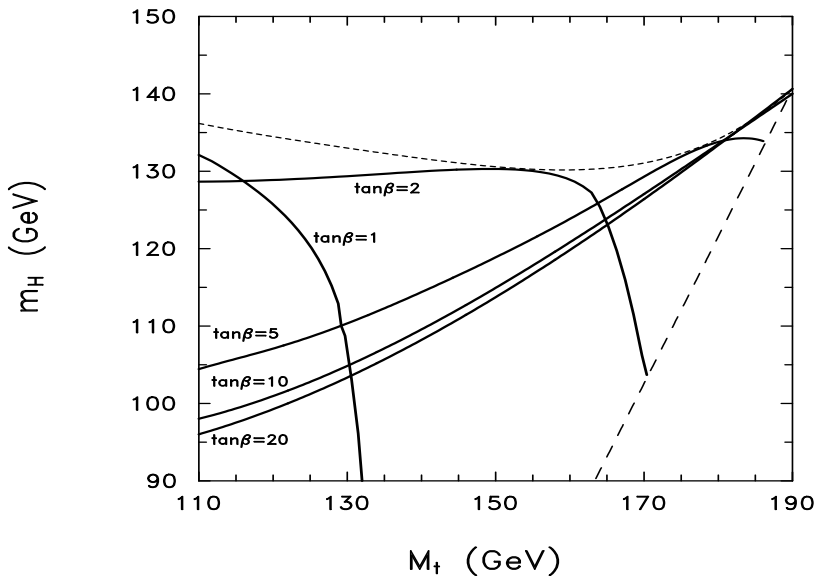,height=7.5cm,width=7cm,bbllx=6.5cm,bblly=3.5cm,bburx=16.cm,bbury=16.5cm}}
\caption{Upper bounds on the lightest scalar Higgs boson in NMSSM with
singlets, different values of $\tan\beta$ and maximal mixing in the
stop sector.}
\label{singlet}
\end{figure}

Assuming that the theory remains perturbative up to the scale
$\Lambda\sim 2\times 10^{16}\ GeV$, 
integrating numerically the RGE and including
radiative corrections~\footnote{Radiative corrections corresponding
to MSSM particles will be retained as in the previous section. We will
assume that those corresponding to non-MSSM couplings are negligible.
For values of the
top-quark mass inside the experimental range the non-MSSM radiative
corrections are found to be small, $\simlt 10$ GeV. 
For a detailed discussion on radiative corrections in these models,
see Refs.~\cite{singlet1,singlet2,singlet3}.} 
for $M_S=1\ TeV$ we find~\cite{10} the upper
bound shown in Fig.~\ref{singlet}  in the $(m_h,m_t)$-plane.

We see from Fig.~\ref{singlet}
that the detailed functional dependence of $m_H$ on $M_t$ is
parametrized by the value of $\tan\beta$.
The dashed curve where the solid lines stop
correspond to values of $m_t$ such that the Yukawa coupling $h_t$
becomes non-perturbative. (For $\tan\beta>30$ the corresponding lines
would follow very close to the 
$\tan\beta=20$ curve in Fig.~\ref{singlet}, but stopping at
lower values of $m_t$ because of the large values of $h_b$.) The dotted
curve on the top of the figure is the enveloping for all values of
$\tan\beta$ and can therefore be considered as the absolute upper
bound. We obtain from Fig.~\ref{singlet}, $m_h\simlt 140$ GeV.

\section{What if a Higgs boson is discovered at FNAL or LEP?}

If the Higgs turns out to be light, it might be discovered at LEP
(for $M_H\simlt 95$ GeV~\cite{LEP2}), or even at FNAL
(for $M_H\simlt 120$ GeV, if Tevatron can run until accumulating
an integrated luminosity $\int{\cal L}\sim$ 25-30 fb$^{-1}$, TeV33 
option~\cite{Marciano}) which could cover all the parameter space
of the MSSM and most of the NMSSM one. In this case, the Higgs
mass measurement could clearly discriminate between the supersymmetric
and non-supersymmetric standard models and, more generally, provide
information about the scale of new physics. These two topics will be
covered in this section.

\subsection{A light Higgs can {\it measure} the scale of New Physics}

\begin{figure}[htb]
\centerline{
\psfig{figure=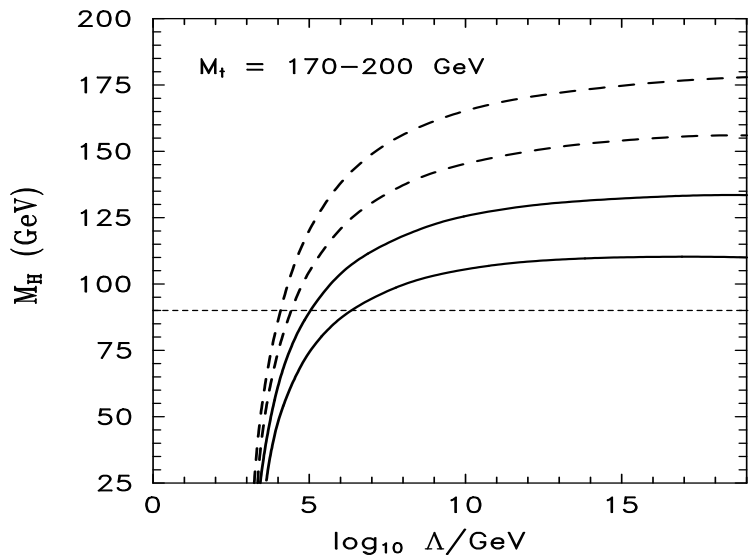,height=7.5cm,width=7cm,bbllx=5.cm,bblly=2.5cm,bburx=14.5cm,bbury=15.5cm}}
\caption{SM lower bounds on $M_H$ from
metastability requirements as a function of 
$\Lambda$ for different values of $M_t$.}
\label{fval6}
\end{figure}

From the bounds on  $M_H(\Lambda)$ previously obtained 
(see Fig.~\ref{fval6}) 
one can easily deduce that
a measurement of $M_H$ will provide an 
{\bf upper bound}  (below the Planck scale) on the
scale of new physics provided that 
\be
\label{final}
M_t[{\rm GeV}]>\frac{M_H[{\rm GeV}]}{2.25}+123\; {\rm GeV}
\ee
%
Thus, the present 
experimental bound from LEP, $M_H>67$ GeV, would imply, from
(\ref{final}), $M_t>153$ GeV, which is fulfilled by experimental 
detection of the top. Even non-observation of the Higgs at 
LEP2 (i.e. $M_H\simgt 95$ GeV), would leave an open window 
($M_t\simgt 165$ GeV) to the possibility that a future Higgs detection
at FNAL or LHC could lead to an upper bound on $\Lambda$. Moreover, Higgs 
detection at LEP2 would put an upper bound on the scale of new physics. 
Taking, for instance,  $M_H\simlt 95$
GeV and  170 GeV $< M_t< $ 180 GeV, then $\Lambda\simlt 10^7$ GeV, while for
180 GeV $< M_t <$ 190 GeV, $\Lambda\simlt 10^4$ 
GeV, as can be deduced from Fig.~\ref{fval6}. Finally, using as upper
bound for the top-quark mass $M_t<181$ GeV we obtain
from (\ref{final}) that only if the condition
\be
M_h>128\ {\rm GeV}
\ee
is fulfilled, the SM can be a consistent theory up to the Planck
scale, where gravitational effects can no longer be neglected.

\subsection{Disentangling between supersymmetric and 
non-supersymmetric models}

We will conclude with a very interesting case, 
where the Higgs sector of the
MSSM plays a key role in the detection of supersymmetry. 
It is the case where all supersymmetric particles are superheavy,
$M_S \sim 1-10\ {\rm TeV}$, and escape detection at LHC. 
In the Higgs sector ${\cal H},A,H^\pm$
decouple, while the $H$ couplings go to the SM $\phi$ couplings, 
$HXY\longrightarrow (\phi XY)_{\rm SM}$
as $\sin(\beta-\alpha)\rightarrow 1$, or are indistinguisable
from the SM ones [see Table 1], $h_u\sin\beta \equiv  h_u^{\rm SM}$,
$h_{d,\ell}\cos\beta  \equiv h_{d,\ell}^{\rm SM}$. 
In this way the $\tan\beta$ dependence of the couplings, either
disappears or is absorbed in the SM couplings.

\begin{figure}[htb]
\centerline{
\psfig{figure=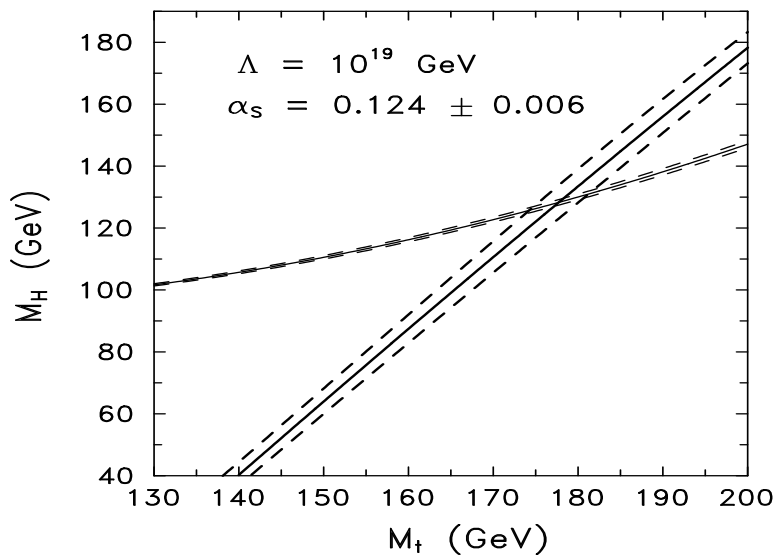,height=7.5cm,width=7cm,bbllx=5.cm,bblly=2.cm,bburx=14.5cm,bbury=15cm}}
\caption{SM lower bounds on $M_H$ (thick lines) as a function of
$M_t$, for $\Lambda=10^{19}$ GeV, from metastability requirements,
and upper bound on the lightest Higgs boson mass in the MSSM
(thin lines) for $M_S=1$ TeV.}
\label{fval5}
\end{figure}

However, from the previous sections it should be clear 
that the Higgs and top mass measurements
could serve to discriminate between the SM and its extensions, 
and to provide information about the
scale of new physics $\Lambda$. 
In Fig.~\ref{fval5}
we give the SM lower bounds on
$M_H$ for $\Lambda\simgt 10^{15}$ GeV (thick lines) and
the upper bound on the mass of the
lightest Higgs boson in the MSSM (thin lines) 
for $M_S\sim 1$ TeV. Taking $M_t=175$ GeV  
and $M_H\simgt 130$ GeV, the SM is
allowed and the MSSM is excluded. On the other hand, 
if $M_H\simlt 130$ GeV, then the MSSM is
allowed while the SM is excluded. Likewise there 
are regions where the SM is excluded, others
where the MSSM is excluded and others where both are permitted or 
both are excluded.

\section{Conclusion}

To conclude, we can say that the search of the Higgs boson at present
and future colliders is, not only an experimental challenge, being the
Higgs boson the last missing ingredient of the Standard Model, but
also a theoretically  appealing question from the 
more fundamental point of view of
physics beyond the Standard Model. In fact, if we are lucky enough and
the Higgs boson is detected soon (preferably at LEP2) and {\it light}, 
its detection
might give sensible information about the possible existence of
new physics. In that case, the
experimental search of the new physics should be urgent and compelling, 
since the existence of new phenomena
might be necessary for our present understanding of physics
for energies at reach in the planned accelerators.

\section*{Acknowledgements}

Work supported in part by the European Union (contract CHRX-CT92-0004) 
and CICYT of Spain (contract AEN95-0195).
I wish to thank my collaborators A.~Brignole, M.~Carena, J.A.~Casas,
J.R.~Espinosa, H.~Haber, G.~Kane, J.~Moreno, A.~Riotto, 
C.~Wagner and F.~Zwirner for what I learnt from them.

\appendixA{Appendix A: $\Delta\Pi$ in the SM}

The quantity $\Delta \Pi$ in the Landau gauge
is given by the sum of the following terms 
\begin{eqnarray}
\Delta \Pi&=&\Delta \Pi_{tt}
\:\:\:\: ({\rm top}\:\:{\rm contribution})\nonumber\\
&+&
\Delta \Pi_{W^{\pm}W^{\mp}}+\Delta \Pi_{Z^{0}Z^{0}}\nonumber\\
&+&\Delta \Pi_{W^{\pm}\chi^{\mp}}+
\Delta \Pi_{Z^{0}\chi_{3}}
\:\:\:\: (\mbox{gauge and Goldstone bosons contribution})\nonumber\\
&+& \Delta \Pi_{\chi^{\pm}\chi^{\mp}}+
\Delta \Pi_{\chi_{3}\chi_{3}}\nonumber\\
&+&\Delta \Pi_{HH}\:\:\:\: (\mbox{pure scalar bosons 
contribution}).
\label{nuevo}
\end{eqnarray}
In Eq.~(\protect\ref{nuevo}) we have 
taken into account only the contribution from the heaviest fermion, the top, 
and indicated by $\chi^{\pm}$ and $\chi_{3}$ the charged and the neutral
Goldstone bosons, respectively.

The complete expression~\cite{CEQR} for the different
contributions to (real part of) 
$\Delta \Pi$ calculated in the $\overline{MS}$ scheme is, for 
$M_H<2 M_W$:
\begin{equation}
\Delta \Pi_{tt}=\frac{3h_{t}^{2}}{8\pi^2}
\left\{-2M_{t}^2\left[Z\left(\frac{M_{t}^2}{M_H^2}\right)
-2\right]+\frac{1}{2}M_H^2\left[\log
\frac{M_{t}^2}{\mu^2}+
Z\left(\frac{M_{t}^2}{M_H^2}\right)
-2\right]\right\}.
\end{equation}

where $Z(x)$ is the function 
\begin{eqnarray}
Z(x)&=&\left\{
\begin{array}{ll}
2 A \:{\rm tan}^{-1}(1/A), & \mbox{if $x>1/4$}\\
A \:\log\left[(1+A)/(1-A)\right],&\mbox{if $x<1/4$}
\end{array}\right.\nonumber\\
A&\equiv&|1-4x|^{1/2}.
\end{eqnarray}

\begin{eqnarray}
&&\Delta \Pi_{W^{\pm}W^{\mp}}+\Delta \Pi
_{Z^{0}Z^{0}} +\Delta \Pi_{W^{\pm}\chi^{\mp}}+
\Delta \Pi_{Z^{0}\chi_{3}}\nonumber\\
&=&\frac{g^2 M_{W}^2}{8\pi^2}
\left[-3 +\frac{5}{4}\frac{M_H^2}{M_W^2} 
+\frac{1}{2}\left(3-\frac{M_H^2}{M_W^2}+\frac{M_H^4}{4 M_W^4}\right)
Z\left(\frac{M_W^2}{M_H^2}\right)
-\frac{M_H^4}{8 M_W^4}\log\frac{M_H^2}{M_W^2}\right.
\nonumber\\
&-&\left.\frac{3M_H^2}{4M_W^2}\log\frac{M_W^2}{\mu^2}+\frac{i\pi}{8}
\frac{M_H^2}{M_W^2}
\right]+\frac{1}{2}
\left\{ \begin{array}{c}
M_{W}\rightarrow M_{Z}\cr\\
g^2\rightarrow g^2+g^{\prime 2}
\end{array}
\right\}.
\end{eqnarray}

The contribution to $\Delta \Pi$
coming from the pure scalar sector deserves more
attention. In the Landau gauge
the Goldstone bosons $\chi$'s do have a field dependent
mass $m_{\chi}(\phi)=-m^2+\lambda \phi^2/2$ which vanishes at the minimum
of the potential $V_{{\rm eff}}(\phi)$. As a consequence, the running mass
$m_{H}$ presents an infrared logarithmic divergence when Goldstone bosons
are included in the effective potential $V_{\rm eff}$, which is 
cancelled by an equal (and opposite in sign) contribution
of the same excitations to $\Delta \Pi$. Explicitly,
\begin{equation}
\label{B15}
\Delta M_{H}^{2}=
\frac{3}{128\pi^2}\frac{g^2 M_H^4}{M_{W}^2}\left[\pi\sqrt{3}
-8+4\log\frac{M_H^2}{m_t^2}\right].
\end{equation}

\appendixB{Appendix B: $\Delta\Pi$ in the MSSM}

We first define
the different self-energy contributions as
\be
\label{sumapol}
\Delta\Pi_{\exis}(M_{\exis}^2)=\Delta\Pi_{\exis}^{(t)}(M_{\exis}^2)
+\Delta\Pi_{\exis}^{(b)}(M_{\exis}^2)
+\Delta\Pi_{\exis}^{(\st\;)}(M_{\exis}^2)
+\Delta\Pi_{\exis}^{(\sb\;)}(M_{\exis}^2)
\ee
where ${\exis}=H,{\cal H},A$. The different contributions in
(\ref{sumapol}) read:
\be
\label{ht}
\Delta\Pi^{(t)}_H(M_H^2)=\frac{3}{8\pi^2}h_t^2\cos^2\alpha
\left[-2M_t^2+\frac{1}{2}M_H^2\right]f(M_t^2,M_t^2,M_H^2)
\ee
\be
\label{Ht}
\Delta\Pi^{(t)}_{\cal H}(M_{\cal H}^2)=\frac{3}{8\pi^2}h_t^2\sin^2\alpha
\left[-2M_t^2+\frac{1}{2}M_{\cal H}^2\right]f(M_t^2,M_t^2,M_{\cal H}^2)
\ee
\be
\label{At}
\Delta\Pi^{(t)}_A(M_A^2)=\frac{3}{8\pi^2}h_t^2\cos^2\beta
\left[-\frac{1}{2}M_A^2\right]f(M_t^2,M_t^2,M_A^2)
\ee
\be
\label{hb}
\Delta\Pi^{(b)}_H(M_H^2)=\frac{3}{8\pi^2}h_b^2\sin^2\alpha
\left[-2m_b^2+\frac{1}{2}M_H^2\right]f(m_b^2,m_b^2,M_H^2)
\ee
\be
\label{Hb}
\Delta\Pi^{(b)}_{\cal H}(M_{\cal H}^2)=\frac{3}{8\pi^2}h_b^2\cos^2\alpha
\left[-2m_b^2+\frac{1}{2}M_{\cal H}^2\right]f(m_b^2,m_b^2,M_{\cal H}^2)
\ee
\be
\label{Ab}
\Delta\Pi^{(b)}_A(M_A^2)=\frac{3}{8\pi^2}h_b^2\sin^2\beta
\left[-\frac{1}{2}M_A^2\right]f(m_b^2,m_b^2,M_A^2)
\ee
\be
\label{xstop}
\Delta\Pi_{\exis}^{(\st\;)}(M_{\exis}^2)=\sum_{i,j=1}^2 
\frac{3}{16\pi^2}\left|C_{{\varphi}ij}^{(\st\;)}\right|^2
f(\msti^2,\mstj^2,M_{\exis}^2)
\ee
\be
\label{xsbottom}
\Delta\Pi_{\exis}^{(\sb\;)}(M_{\exis}^2)=\sum_{i,j=1}^2 
\frac{3}{16\pi^2}\left|C_{{\varphi}ij}^{(\sb\;)}\right|^2
f(\msbi^2,\msbj^2,M_{\exis}^2)
\ee
The different coefficients in (\ref{xstop}) and (\ref{xsbottom})
are:
\bea
\label{hstop}
C_{Hij}^{(\st\;)} & = &
\frac{2 \sqrt{2} \sin^2\theta_W}{3} \frac{M_Z^2}{v}
\sin(\beta+\alpha) \left[ \delta_{ij} + \frac{ 3 - 8
\sin^2\theta_W}{4 \sin^2\theta_W} Z_U^{1i} Z_U^{1j} \right]
\\
& - & \sqrt{2} h_t^2v\sin\beta\cos\alpha \ \delta_{ij}
-\frac{1}{\sqrt{2}}h_t(A_t\cos\alpha+\mu\sin\alpha)
(Z_U^{1i *}Z_U^{2j}+Z_U^{1j}Z_U^{2i *}) \nonumber
\eea
\bea
\label{Hstop}
C_{{\cal H}ij}^{(\st\;)}&=&
-\frac{2 \sqrt{2} \sin^2\theta_W}{3} \frac{M_Z^2}{v}
\cos(\beta+\alpha) \left[ \delta_{ij} + \frac{ 3 - 8
\sin^2\theta_W}{4 \sin^2\theta_W} Z_U^{1i} Z_U^{1j} \right]
\\
&-&\sqrt{2}h_t^2 v\sin\beta\sin\alpha \ \delta_{ij}
-\frac{1}{\sqrt{2}}h_t(A_t\sin\alpha-\mu\cos\alpha)
(Z_U^{1i *}Z_U^{2j}+Z_U^{1j}Z_U^{2i *})\nonumber
\eea
\be
\label{Astop}
C_{Aij}^{(\st\;)}=
-\frac{1}{\sqrt{2}}h_t(A_t\cos\beta+\mu\sin\beta)
(Z_U^{1i *}Z_U^{2j}-Z_U^{1j}Z_U^{2i *})
\ee
\bea
\label{hsbottom}
C_{Hij}^{(\sb\;)}&=&
-\frac{ \sqrt{2} \sin^2\theta_W}{3} \frac{M_Z^2}{v}
\sin(\beta+\alpha) \left[ \delta_{ij} + \frac{ 3 - 4
\sin^2\theta_W}{2 \sin^2\theta_W} Z_D^{1i} Z_D^{1j} \right]
\\
& + & \sqrt{2}h_b^2v\cos\beta\sin\alpha \ \delta_{ij}
+\frac{1}{\sqrt{2}}h_b(A_b\sin\alpha+\mu\cos\alpha)
(Z_D^{1i *}Z_D^{2j}+Z_D^{1j}Z_D^{2i *})\nonumber
\eea
\bea
\label{Hsbottom}
C_{{\cal H}ij}^{(\sb\;)}&=&
\frac{ \sqrt{2} \sin^2\theta_W}{3} \frac{M_Z^2}{v}
\cos(\beta+\alpha) \left[ \delta_{ij} + \frac{ 3 - 4
\sin^2\theta_W}{2 \sin^2\theta_W} Z_D^{1i} Z_D^{1j} \right]
\\
& - & \sqrt{2}h_b^2v\cos\beta\cos\alpha \ \delta_{ij}
-\frac{1}{\sqrt{2}}h_b(A_b\cos\alpha-\mu\sin\alpha)
(Z_D^{1i *}Z_D^{2j}+Z_D^{1j}Z_D^{2i *}) \nonumber
\eea
\be
\label{Asbottom}
C_{Aij}^{(\sb\;)}=
\frac{1}{\sqrt{2}}h_b(A_b\sin\beta+\mu\cos\beta)
(Z_D^{1i *}Z_D^{2j}-Z_D^{1j}Z_D^{2i *}),
\ee
where the matrices $Z_U^{ij}$ and $Z_D^{ij}$ are those diagonalizing the
stop and sbottom squared mass matrices, respectively, and the function 
$f(m_1^2,m_2^2,q^2)$ which arises from the integration of the
loop of (scalar) particles, is given by
$$
f(m_1^2,m_2^2,q^2)=-1+\frac{1}{2}
\left(\frac{m_1^2+m_2^2}{m_1^2-m_2^2}-\ \delta\right)
\log\frac{m_2^2}{m_1^2}+\frac{1}{2}r\log\left[\frac{(1+r)^2-\ \delta^2}
{(1-r)^2-\ \delta^2}\right]
$$
with $\delta=(m_1^2-m_2^2)/q^2$ and
${\displaystyle r=\sqrt{(1+\delta)^2-\frac{4m_1^2}{q^2}} }$


\end{document}